\documentclass{amsart}[11pt]

\theoremstyle{plain}
\newtheorem{theorem}{Theorem}[section]

\newtheorem{proposition}[theorem]{Proposition}
\newtheorem{definition}[theorem]{Definition}

\theoremstyle{remark}
\newtheorem{remark}[theorem]{Remark}
\numberwithin{equation}{section}

\input{tcilatex}

\begin{document}
\thispagestyle{empty}
\begin{center}
{\footnotesize Available at: 
{\tt http://publications.ictp.it}}\hfill IC/2007/108\\
\vspace{1cm}
United Nations Educational, Scientific and Cultural Organization\\
and\\
International Atomic Energy Agency\\
\medskip
THE ABDUS SALAM INTERNATIONAL CENTRE FOR THEORETICAL PHYSICS\\
\vspace{1.8cm}
{\bf INTRODUCTIVE BACKGROUNDS OF MODERN QUANTUM MATHEMATICS\\
WITH APPLICATION TO NONLINEAR DYNAMICAL SYSTEMS}\\
\bigskip
{\small{\it The authors dedicate this article to their friend and teacher
academician Prof. Anatoliy M. Samoilenko on the occasion of his 70th
Birthday with great compliments and gratitude to his brilliant
talent and impressive impact to modern theory of nonlinear dynamical systems
of mathematical physics and nonlinear analysis.}}\\

\vspace{1.5cm}
Anatoliy K. Prykarpatsky\footnote{pryk.anat@ua.fm, prykanat@cybergal.com}\\
{\it The AGH University of Science and Technology, Krak\'{o}w 30-059, Poland,\\
National Academy of Sciences of Ukraine, Lviv, Ukraine \\
and \\
The Abdus Salam International Centre for Theoretical Physics, Trieste, Italy,}\\[1em]
Nikolai N. Bogoliubov Jr.\footnote{nikolai\_bogolubov@hotmail.com}\\
{\it V.A. Steklov Mathematical Institute of RAN, Moscow, Russian Federation \\
and \\
The Abdus Salam International Centre for Theoretical Physics, Trieste, Italy,}\\[1em]
Jolanta Golenia\footnote{goljols@tlen.pl}\\
{\it Department of Applied Mathematics, AGH University of Science and Technology,\\ Krak\'{o}w 30-059, Poland}\\[1em]
and\\[1em]
Ufuk Taneri\footnote{ufuk.taneri@gmail.com}\\
{\it Department of Applied Mathematics and Computer Science, Eastern Mediterranean University EMU, Famagusta, North Cyprus \\
and\\
Kyrenia American University GAU, Institute of Graduate Studies, Kyrenia, North
Cyprus.}\\

\end{center}
\baselineskip=18pt

\vfill
\begin{center}
MIRAMARE -- TRIESTE\\
September 2007\\
\end{center}
\vfill
\newpage
\setcounter{page}{1}
\centerline{\bf Abstract}
\medskip

Introductive backgrounds of a new mathematical physics discipline - Quantum
Mathematics - are discussed and analyzed both from historical and analytical
points of view. The magic properties of the second quantization method,
invented by V. Fock in 1934, are demonstrated, and an impressive
application to the nonlinear dynamical systems theory is considered.
\newpage

\section{INTRODUCTION}

There is a broad and inclusive view of modern mathematical physics by many
mathematicians and mathematical physicists. During the last century, modern
mathematical physics evolved within at least four components which
illustrate \cite{JQ} the development of the mathematics and quantum physics
synergy:

1) the use of ideas from mathematics in shedding new light on the existing
principles of quantum physics, either from a conceptual or from a quantitive
point of view;\-

2) the use of ideas from mathematics in discovering new "laws of quantum
physics";\-

3) the use of ideas from quantum physics in shedding new light on existing
mathematical structures;

4) the use of ideas from quantum physics in discovering new domains in
mathematics.

Each one of these topics plays some role in understanding the modern
mathematical physics. However, our success in directions 2) and 4) is
certainly more modest than our success in directions 1) and 3). In some
cases it is difficult to draw a clear-cut distinction between these two
sets. In fact, we are lucky when it is possible to make progress in
directions 2) and 4); so much so that when we achieve a major progress,
historians like to speak of a revolution. In any case, many of mathematical
physicists strive to understand within their research efforts these deep and
lofty goals. There are many situations howevere, when mathematical
physicists' research efforts are directed toward one other more mundane
aspect:

5) the use of ideas from quantum physics and mathematics to benefit
"economic competitiveness".

Here too, one might subdivide this aspect into conceptual understanding on
one hand (such as the mathematical model of Black and Sholes for pricing of
derivative securities in financial markets) and invention on the other: the
formulation of new algorithms or materials (e.g. quantum computers) which
might revolutionize technology or change our way of life. As in the first
four cases, the boundary between these domains is not sharp, and it remains
open to views and interpretations. This fifth string can be characterized as
"applied" mathematical physics. We will restrict our analysis to the first
four strands characterizing modern quantum physics and mathematics aspects;
it is believed that that most of the profound applied directions arise after
earlier fundamental quantum physics and mathematics progress.

We have passed through an extraordinary 35-year period of development of
modern fundamental mathematics and quantum physics. Much of this development
has drawn from one subject to understand the other. Not only concepts from
diverse fields have been united: statistical physics, quantum field theory
and functional integration; gauge theory and geometry; index theory and knot
invariants, etc., but also, new phenomena have been recognized and new areas
have emerged whose significance for both mathematics and modern quantum
physics is only partially understood: for example, non-commutative geometry,
super-analysis, mirror symmetry, new topological invariants of manifolds,
and the general notion of geometric quantization.

There is no doubt that, over the past 35 years, the ideas from quantum
physics have led to far greater inventions of new mathematics than the ideas
from mathematics have in discoveries of laws of quantum physics. Recognition
of this underlines the opportunities for future progress in the opposite
direction: a new understanding of the quantum nature of the world is
certainly our expectation!

Great publicity and recognition has been attached to the progress made in
modern geometry, representation theory, and deformation theory due to this
interaction. But one should ignore the substantial progress in analysis and
probability theory, which unfortunately is more difficult to understand
because of its delicate dependence on subtle notions of continuity.

On the other hand, there are deep differences between pure mathematics and
modern quantum physics fundamentals. They have evolved from different
cultures and they each have a distinctive set of values of their own, suited
for their different realms of universality. But both subjects are strongly
based on \textit{\ intuition}, some natural and some acquired, which form
our understanding. Quantum physics describes the natural micro-world. Hence,
physicists appeal to observation in order to verify the validity of a
physical theory. And, although much of mathematics arises from the natural
world, mathematics has no analogous testing grounds - mathematicians appeal
to their own set of values, namely mathematical proof, to justify validity
of a mathematical theory. In mathematical physics, when announcing results
of a mathematical nature, it is necessary to claim a theorem when \ the
proof meets the mathematical community standards for a proof; otherwise, it
is necessary to make a conjecture with a detailed outline for support. Most
of physics, on the other hand, has completely different standards.

There is no question that the interaction between modern mathematics and
quantum physics will change radically during this running century. We do
hope however, that this evolution will preserve the positive experience of
being a mathematician, a pure physicist, or a mathematical physicist, so
that it remains attractive to the brightest and gifted young students today
and tomorrow.

It is instructive to look at the beginning of the \emph{XX }$^{th}$ century
and trace the way mathematics has been exerting influence on modern and
classical quantum physics, and next observe the way the modern quantum
physics is nowadays exerting so impressive influence on modern mathematics.
With the latter,  application of modern quantum mathematics to studying
nonlinear dynamical systems in functional spaces will for example be a
significant topic of our present work. We will begin with a brief history of
quantum mathematics: 

\emph{The beginning of \ the XX }$^{th}$ \emph{century:} 

\begin{itemize}
\item P.A.M. Dirac -- first realized and used the fact that the commutator
operation $\ D_{a}:\mathcal{A}\ni b\longrightarrow \lbrack a,b]\in \mathcal{A%
},$ \ where $a\in \mathcal{A}$ is fixed and $b\in \mathcal{A},$ is a
differentiation of any operator algebra $\mathcal{A}$; moreover, he first
constructed a spinor matrix realization of the Poincar\'{e} symmetry group $%
\mathcal{P}(1,3),$ \cite{Dir} (1920-1926);

\item J. von Neumann -- first applied the spectral theory of self-adjoint
operators in Hilbert spaces to explain the radiation spectra of atoms and
the stability of the related matter, \cite{Ne} (1926);

\item V. Fock -- first introduced the notion of many-particle Hilbert space,
Fock space, and introduced the related creation and annihilation operators
acting in it, \cite{Fo} (1932);

\item H. Weyl -- first understood the fundamental role of the notion of
symmetry in physics and developed a physics-oriented group theory; moreover
he showed the importance of different representations of classical matrix
groups for physics and studied the unitary representations of the
Heisenberg-Weyl group related with creation and annihilation operators in
Fock space, \cite{We} (1931).
\end{itemize}

\bigskip

\emph{The end of the\ XX }$^{th}$ \emph{century:} \medskip

New developments are due to

\begin{itemize}
\item L. Faddeev with co-workers -- quantum inverse spectral theory
transform, \cite{Fa} (1978);

\item V. Drinfeld, S. Donaldson, E. Witten -- quantum groups and algebras,
quantum topology, quantum super-analysis, \cite{Dr,Do,Wi} (1982-1994);

\item Yu. Manin, R. Feynman -- quantum information theory, \cite{Ma,Fe,Fe1}
(1980-1986);

\item P. Shor, E. Deutsch, L. Grover and others -- quantum computer
algorithms, \cite{Sh,De,Gr} (1985-1997).
\end{itemize}

As one can observe, many exciting and highly important mathematical
achievements were strictly motivated by the impressive and deep influence of
quantum physics ideas and ways of thinking, leading nowadays to an
altogether new scientific field often called quantum mathematics.

Following this quantum mathematical way of thinking, we will demonstrate
below that a wide class of strictly nonlinear dynamical systems in
functional spaces can be treated as a natural object in specially
constructed Fock spaces in which the corresponding evolution flows are
completely linearized. Thereby, the powerful machinery of classical
mathematical tools can be applied to studying the analytical properties of
exact solutions to suitably well posed Cauchy problems.

\section{ Mathematical preliminaries: Fock space and its realizations}

Let $\Phi $ be a separable Hilbert space, $F$ be a topological real linear
space and $\mathcal{A}:=\left\{ A(\varphi ):\varphi \in F\right\} $ a family
of commuting self-adjoint operators in $\Phi $ (i.e. these operators commute
in the sense of their resolutions of the identity). Consider the Gelfand
rigging \cite{Be} of the Hilbert space ${\Phi ,}$ i.e., a chain 
\begin{equation}
\mathcal{D}\subset {\Phi }_{+}\subset {\Phi }\subset {\Phi }_{-}\subset 
\mathcal{D^{^{\prime }}}  \label{eq2.1}
\end{equation}%
in which ${\Phi }_{+}$ and ${\Phi }_{-}$ are further Hilbert spaces, and the
inclusions are dense and continuous, i.e. ${\Phi }_{+}$ is topologically
(densely and continuously) and quasi-nuclearly (the inclusion operator $i:{%
\Phi }_{+}\longrightarrow {\Phi }$ is of the Hilbert - Schmidt type)
embedded into ${\Phi }$, ${\Phi }_{-}$ is the dual of ${\Phi }_{+}$ with
respect to the scalar product $<.,.>_{{\Phi }}$ in ${\Phi }$, and $\mathcal{D%
}$ is a separable projective limit of Hilbert spaces, topologically embedded
into ${\Phi }_{+}$. Then, the following structural theorem \cite{Be,BK}
holds:

\begin{theorem}
\label{th2.1} Assume that the family of operators $\mathcal{A}$ \ satisfies
the following conditions:

a)$\quad \mathcal{D}\subset DomA(\varphi ),\;\varphi \in F,$ and the closure
of the operator $A(\varphi )\uparrow \mathcal{D}$ coincides with $A(\varphi
) $ for any $\varphi \in F$, that is $A(\varphi )\uparrow \mathcal{D}%
=A(\varphi )$ in ${\Phi }$;

b) the Range $A(\varphi )\uparrow \mathcal{D}\subset {\Phi }_{+}$ \ for any $%
\varphi \in F$;

c) for every $f\in \mathcal{D}$ the mapping $F\ni \varphi \longrightarrow
A(\varphi )f\in {\Phi }_{+}$ is linear and continuous;

d) there exists a strong cyclic (vacuum) vector $|\Omega \rangle \in
\bigcap_{\varphi \in F}DomA(\varphi ),$ such that the set of all vectors $%
|\Omega \rangle ,$ $\prod_{j=1}^{n}A(\varphi _{j})|\Omega \rangle ,$ $n\in 
\mathbb{Z}_{+},$ is total in ${\Phi }_{+}$ (i.e. their linear hull is dense
in ${\Phi }_{+}$).

Then there exists a probability measure $\mu $ on $(F^{\prime },C_{\sigma
}(F^{\prime }))$, where $F^{\prime }$ is the dual of $F$ and $C_{\sigma
}(F^{\prime })$ is the $\sigma -$algebra generated by cylinder sets in $%
F^{\prime }$ such that, for $\mu -$almost every $\eta \in F^{\prime }$ there
is a generalized joint eigenvector $\omega (\eta )\in {\Phi }_{-}$ of the
family $\mathcal{A}$, corresponding to the joint eigenvalue $\eta \in
F^{\prime }$, that is 
\begin{equation}
<\omega (\eta ),A(\varphi )f>_{{\Phi }}=\eta (\varphi )<\omega(\eta ),f>_{{%
\Phi }}  \label{eq2.1a}
\end{equation}%
with $\eta (\varphi )\in \mathbb{R}$ denoting the pairing between $F$ and $%
F^{\prime }$.

The mapping 
\begin{equation}
{\Phi }_{+}\ni f\longrightarrow <\omega (\eta ),f>_{{\Phi }}:=\hat{f}(\eta
)\in \mathbb{C}  \label{eq2.1b}
\end{equation}%
for any $\eta \in F^{\prime }$ can be continuously extended to a unitary
surjective operator $\mathcal{F}:{\Phi }\longrightarrow L_{2}^{(\mu
)}(F^{\prime };\mathbb{C}),$ where 
\begin{equation}
\mathcal{F}f(\eta ):=\hat{f}(\eta )  \label{eq2.1c}
\end{equation}%
for any $\eta \in F^{\prime }$ is a generalized Fourier transform,
corresponding to the family $\mathcal{A}$. Moreover, the image of the
operator $A(\varphi )$, $\varphi \in F^{\prime }$, under the $\mathcal{F}-$%
mapping is the operator of multiplication by the function $F^{\prime }\ni
\eta \rightarrow \eta (\varphi )\in \mathbb{C}$.
\end{theorem}

We assume additionally that the main Hilbert space $\Phi $ possesses the
standard Fock space (bose)-structure \cite{BB,Ber,PTB}, that is 
\begin{equation}
{\Phi }=\oplus _{n\in \mathbb{Z}_{+}}{\Phi }_{n},  \label{eq2.1'}
\end{equation}%
where subspaces ${\Phi }_{n}:={\Phi }_{(s)}^{\otimes n},$ $n\in \mathbb{Z}%
_{+}$, are the symmetrized tensor products of a Hilbert space $\mathcal{H}%
:=L_{2}(\mathbb{R}^{m};\mathbb{C}).$ If a vector $%
g:=(g_{0},g_{1},...,g_{n},...)\in \Phi $, its norm 
\begin{equation}
\Vert g\Vert _{\Phi }:=\left( \sum_{n\in \mathbb{Z}_{+}}\Vert g_{n}\Vert
_{n}^{2}\right) ^{1/2},  \label{eq2.2}
\end{equation}%
where $g_{n}\in {\Phi }_{(s)}^{\otimes n}\simeq L_{2,(s)}((\mathbb{R}%
^{m})^{n};\mathbb{C})$ and $\parallel ...\parallel _{n}$ is the
corresponding norm in ${\Phi }_{(s)}^{\otimes n}$ for all $n\in \mathbb{Z}%
_{+}$. Denote here that, concerning the rigging structure (\ref{eq2.1}),
there holds the corresponding rigging for the Hilbert spaces ${\Phi }%
_{(s)}^{\otimes n}$, $n\in \mathbb{Z}_{+}$, that is 
\begin{equation}
\mathcal{D}_{(s)}^{n}\subset {\Phi }_{(s),+}^{\otimes n}\subset {\Phi }%
_{(s)}^{\otimes n}\subset {\Phi }_{(s),-}^{\otimes n}  \label{eq2.3}
\end{equation}%
with some suitably chosen dense and separable topological spaces of
symmetric functions $\mathcal{D}_{(s)}^{n}$, $n\in \mathbb{Z}_{+}$.
Concerning expansion (\ref{eq2.1'}) we obtain by means of projective and
inductive limits \cite{Be,Ber,BK} the quasi-nucleous rigging of the Fock
space $\Phi $ in the form (\ref{eq2.1}):%
\begin{equation*}
\mathcal{D}\subset {\Phi }_{+}\subset {\Phi }\subset {\Phi }_{-}\subset 
\mathcal{D^{^{\prime }}}.
\end{equation*}

Consider now any vector $|(\alpha )_{n}\rangle \in {\Phi }_{(s),}^{\otimes
n} $ $n\in \mathbb{Z}_{+}$, which can be written \cite{Be,BB,KS} in the
following canonical Dirac ket-form: 
\begin{equation}
|(\alpha )_{n}\rangle :=|\alpha _{1},\alpha _{2},...,\alpha _{n}\rangle ,
\label{eq2.4}
\end{equation}%
where, by definition, \ 
\begin{equation}
|\alpha _{1},\alpha _{2},...,\alpha _{n}\rangle :=\frac{1}{\sqrt{n!}}%
\sum_{\sigma \in S_{n}}|\alpha _{\sigma (1)}\rangle \otimes |\alpha _{\sigma
(2)}\rangle ...|\alpha _{\sigma (n)}\rangle  \label{eq2.5}
\end{equation}%
and $|\alpha _{j}\rangle \in {\Phi }_{(s)}^{\otimes 1}(\mathbb{R}^{m};%
\mathbb{C}):=\mathcal{H}$ for any fixed $j\in \mathbb{Z}_{+}$. The
corresponding scalar product of base vectors as (\ref{eq2.5}) is given as
follows: 
\begin{equation}
\begin{array}{c}
\langle (\beta )_{n}|(\alpha )_{n}\rangle :=\langle \beta _{n},\beta
_{n-1},...,\beta _{2},\beta _{1}|\alpha _{1},\alpha _{2},...,\alpha
_{n-1},\alpha _{n}\rangle \\[5pt] 
=\sum_{\sigma \in S_{n}}\langle \beta _{1}|\alpha _{\sigma (1)}\rangle
...\langle \beta _{n}|\alpha _{\sigma (n)}\rangle :=per\{\langle \beta
_{i}|\alpha _{j}\rangle :i,j=\overline{1,n}\},%
\end{array}
\label{eq2.6}
\end{equation}%
where $"per"$ denotes the permanent of matrix and $\langle .|.\rangle $ is
the corresponding product in the Hilbert space $\mathcal{H}$. Based now on
representation (\ref{eq2.4}) one can define an operator $a^{+}(\alpha ):{%
\Phi }_{(s)}^{\otimes n}\longrightarrow {\Phi }_{(s)}^{\otimes (n+1)}$ for
any $|\alpha \rangle \in \mathcal{H}$ as follows: 
\begin{equation}
a^{+}(\alpha )|\alpha _{1},\alpha _{2},...,\alpha _{n}\rangle :=|\alpha
,\alpha _{1},\alpha _{2},...,\alpha _{n}\rangle ,  \label{eq2.7}
\end{equation}%
which is called the "creation" operator in the Fock space $\Phi $. The
adjoint operator $a(\beta ):=(a^{+}(\beta ))^{\ast }:{\Phi }_{(s)}^{\otimes
(n+1)}\longrightarrow {\Phi }_{(s)}^{\otimes n}$ with respect to the Fock
space ${\Phi }$ (\ref{eq2.1'}) for any $|\beta \rangle \in \mathcal{H},$
called the "annihilation" operator, acts as follows: 
\begin{equation}
a(\beta )|\alpha _{1},\alpha _{2},...,\alpha _{n+1}\rangle :=\sum_{\sigma
\in S_{n}}\langle \beta ,\alpha _{j}\rangle |\alpha _{1},\alpha
_{2},...,\alpha _{j-1},\hat{\alpha}_{j},\alpha _{j+1},...,\alpha
_{n+1}\rangle ,  \label{eq2.8}
\end{equation}%
where the $"hat"$ over a vector denotes that it should be omitted from the
sequence.

It is easy to check that the commutator relationship 
\begin{equation}
\lbrack a^{+}(\alpha ),a(\beta )]=\langle \alpha ,\beta \rangle 
\label{eq2.9}
\end{equation}%
holds for any vectors $|\alpha \rangle \in \mathcal{H}$ and $|\beta \rangle
\in \mathcal{H}$. Expression (\ref{eq2.9}), owing to the rigged structure (%
\ref{eq2.1}), can be naturally extended to the general case, when vectors $\
|\alpha \rangle $ and $|\beta \rangle \in \mathcal{H}_{-}$, conserving its
form. In particular, taking $|\alpha \rangle :=|\alpha (x)\rangle =\frac{1}{%
\sqrt{2\pi }}e^{i\langle \lambda ,x\rangle }\in \mathcal{H}_{-}:=L_{2,-}(%
\mathbb{R}^{m};\mathbb{C})$ for any $x\in \mathbb{R}^{m}$, one easily gets
from (\ref{eq2.9}) that 
\begin{equation}
\lbrack a^{+}(x),a(y)]=\delta (x-y),  \label{eq2.10}
\end{equation}%
where we put, by definition, $a^{+}(x):=a^{+}(\alpha (x))$ and $%
a(y):=a(\alpha (y))$ for all $x,y\in \mathbb{R}^{m}$ and denoted by $\delta
(\cdot )$ the classical Dirac delta-function.

The construction above makes it possible to observe easily that there exists
a unique vacuum vector $|\Omega \rangle \in \mathcal{H}_{+}$, such that for
any $x\in \mathbb{R}^{m}$ 
\begin{equation}
a(x)|\Omega \rangle =0,  \label{eq2.11}
\end{equation}%
and the set of vectors 
\begin{equation}
\left( \prod_{j=1}^{n}a^{+}(x_{j})\right) |\Omega \rangle \in {\Phi }%
_{(s)}^{\otimes n}  \label{eq2.12}
\end{equation}%
is total in ${\Phi }_{(s)}^{\otimes n}$, that is their linear integral hull
over the dual functional spaces $\hat{\Phi}_{(s)}^{\otimes n}$ is dense in
the Hilbert space ${\Phi }_{(s)}^{\otimes n}$ for every $n\in \mathbb{Z}_{+}$%
. This means that for any vector $g\in \Phi $ the following representation 
\begin{equation}
g=\oplus _{n\in \mathbb{Z}_{+}}\int_{(\mathbb{R}^{m})^{n}}\hat{g}%
_{n}(x_{1},...,x_{n})a^{+}(x_{1})a^{+}(x_{2})...a^{+}(x_{n})|\Omega \rangle 
\label{eq2.13}
\end{equation}%
holds with the Fourier type coefficients $\hat{g}_{n}\in \hat{\Phi}_{n}:=%
\hat{\Phi}_{(s)}^{\otimes n}$ for all $n\in \mathbb{Z}_{+}$, with $\hat{\Phi}%
_{(s)}^{\otimes 1}:=\mathcal{H}\simeq {L}_{2}(\mathbb{R}^{m};\mathbb{C}).$
The latter is naturally endowed with the dual to (\ref{eq2.1}) Gelfand type
quasi-nucleous rigging 
\begin{equation}
\mathcal{H}_{+}\subset \mathcal{H}\subset \mathcal{H},  \label{eq2.14}
\end{equation}%
making it possible to construct a quasi-nucleous rigging of the dual Fock
space ${\hat{\Phi}}:=\oplus _{n\in \mathbb{Z}_{+}}{\hat{\Phi}}_{n}.$
Thereby, chain (\ref{eq2.14}) generates the dual Fock space quasi-nucleous
rigging 
\begin{equation}
\mathcal{\hat{D}}\subset {\hat{\Phi}}_{+}\subset {\hat{\Phi}}\subset {\hat{%
\Phi}}_{-}\subset \mathcal{\hat{D}}^{\prime }\mathcal{^{^{\prime }}}
\label{eq2.15}
\end{equation}%
with respect to the central Fock type Hilbert space ${\hat{\Phi}},$ where $%
\mathcal{\hat{D}}\simeq \mathcal{D},$ easily following from (\ref{eq2.1})
and (\ref{eq2.14}).

Construct now the following self-adjoint operator 
\begin{equation}
a^{+}(x)a(x):=\rho (x):\Phi \rightarrow \Phi ,  \label{eq2.16}
\end{equation}%
called the density operator at a point $x\in \mathbb{R}^{m},$ satisfying the
commutation properties: 
\begin{equation}
\begin{array}{c}
\lbrack \rho (x),\rho (y)]=0, \\[5pt] 
\lbrack \rho (x),a(y)]=-a(y)\delta (x-y), \\[5pt] 
\lbrack \rho (x),a^{+}(y)]=a^{+}(y)\delta (x-y)%
\end{array}
\label{eq2.17}
\end{equation}%
for all $y\in \mathbb{R}^{m}$.

Now, if to construct the following self-adjoint family $\mathcal{A}:=\left\{
\int_{\mathbb{R}^{m}}\rho (x)\varphi (x)dx:\varphi \in F\right\} $ of linear
operators in the Fock space $\Phi $,\ where $F$ $:=\mathcal{S}(\mathbb{R}%
^{m};\mathbb{R})$ is the Schwartz functional space, one can derive, making
use of Theorem \ref{th2.1}, that there exists the generalized Fourier
transform (\ref{eq2.1c}), such that 
\begin{equation}
{\Phi }(\mathcal{H})=L_{2}^{(\mu )}(\mathcal{S}^{\prime };\mathbb{C})\simeq
\int_{\mathcal{S}^{\prime }}^{\oplus }\Phi _{\eta }d\mu (\eta )
\label{eq2.17a}
\end{equation}%
for some Hilbert space sets $\Phi _{\eta },$ $\eta \in F^{\prime },$ \ and a
suitable measure $\mu $ on $\mathcal{S}^{\prime },$ \ with respect to which
the corresponding joint eigenvector $\omega (\eta )\in \Phi _{+}$ for any $%
\eta \in F^{\prime }$ generates the Fourier transformed family $\hat{%
\mathcal{A}}=\left\{ \eta (\varphi )\in \mathbb{R}:\;\;\varphi \in \mathcal{S%
}\right\} $. Moreover, if $\dim \Phi _{\eta }=1$ for all $\ \eta \in F,$ the
Fourier transformed eigenvector $\hat{\omega}(\eta ):=\Omega (\eta )=1$ for
all $\eta \in F^{^{\prime }}$.

Now we will consider the family of self-adjoint operators $\mathcal{A}$ as
generating a unitary family $\mathcal{U}:=\left\{ U(\varphi ):\varphi \in
F\right\} =\exp (i\mathcal{A}),$ where for any $\rho (\varphi )\in \mathcal{A%
}$, $\varphi \in F$, the operator 
\begin{equation}
U(\varphi ):=\exp [i\rho (\varphi )]  \label{eq2.18}
\end{equation}%
is unitary, satisfying the abelian commutation condition 
\begin{equation}
U(\varphi _{1})U(\varphi _{2})=U(\varphi _{1}+\varphi _{2})  \label{eq2.19}
\end{equation}%
for any $\varphi _{1},\varphi _{2}\in F$.

Since, in general, the unitary family $\mathcal{U}=\exp (i\mathcal{A})$ is
defined in some Hilbert space $\Phi $, not necessarily being of Fock type,
the important problem of describing its Hilbertian cyclic representation
spaces arises, within which the factorization 
\begin{equation}
\rho (\varphi )=\int_{\mathbb{R}^{m}}a^{+}(x)a(x)\varphi (x)dx
\label{eq2.20}
\end{equation}%
jointly with relationships (\ref{eq2.17}) hold for any $\varphi \in F$. This
problem can be treated using mathematical tools devised both within the
representation theory of $C^{\ast }-$algebras \cite{Di} and the
Gelfand--Vilenkin \cite{GV} approach. Below we will describe the main
features of the Gelfand--Vilenkin formalism, being much more suitable for
the task, providing a reasonably unified framework of constructing the
corresponding representations.

\begin{definition}
\label{def2.1} Let $F$ be a locally convex topological vector space, $%
F_{0}\subset F$ be a finite dimensional subspace of $F$. \ Let $%
F^{0}\subseteq F^{\prime }$ be defined by 
\begin{equation}
F^{0}:=\left\{ \xi \in F^{\prime }:\;\;\xi |_{F_{0}}=0\right\} ,
\label{eq2.21}
\end{equation}%
and called the annihilator of $F_{0}$.
\end{definition}

The quotient space $F^{\prime 0}:=F^{\prime }/F^{0}$ may be identified with $%
F_{0}^{\prime }\subset F^{\prime }$, the adjoint space of $F_{0}$.

\begin{definition}
\label{def2.2} Let $A\subseteq F^{\prime }$; then the subset

\begin{equation}
X_{F^{0}}^{(A)}:=\left\{ \xi \in F^{\prime }:\xi +F^{0}\subset A\right\}
\label{eq2.22}
\end{equation}%
is called the cylinder set with base $A$ and generating subspace $F^{0}$.
\end{definition}

\begin{definition}
\label{def2.3} Let $n=\dim F_{0}=\dim F_{0}^{\prime }=\dim F^{\prime 0}$.
One says that a cylinder set $X^{(A)}$ has Borel base, if $A$ is Borel, when
regarded as a subset of $\mathbb{R}^{n}$.
\end{definition}

The family of cylinder sets with Borel base forms an algebra of sets.

\begin{definition}
\label{def2.4} The measurable sets in $F^{\prime }$ are the elements of the $%
\sigma -$ algebra generated by the cylinder sets with Borel base.
\end{definition}

\begin{definition}
\label{def2.5} A cylindrical measure in $F^{\prime }$ is a real-valued $%
\sigma -$pre-additive function $\mu $ defined on the algebra of cylinder
sets with Borel base and satisfying the conditions $0\leq \mu (X)\leq 1$ for
any $X$, $\mu (F^{\prime })=1$ and $\mu \left( \coprod_{j\in \mathbb{Z}%
_{+}}X_{j}\right) =\sum_{j\in \mathbb{Z}_{+}}\mu (X_{j}),$ if all sets $%
X_{j}\subset F^{\prime }$, $j\in \mathbb{Z}_{+}$, have a common generating
subspace $F_{0}\subset F$.
\end{definition}

\begin{definition}
\label{def2.6} A cylindrical measure $\mu $ satisfies the commutativity
condition if and only if for any bounded continuous function $\alpha :%
\mathbb{R}^{n}\longrightarrow \mathbb{R}$ of $n\in \mathbb{Z}_{+}$ real
variables the function 
\begin{equation}
\alpha \lbrack \varphi _{1},\varphi _{2},...,\varphi _{n}]:=\int_{F^{\prime
}}\alpha (\eta (\varphi _{1}),\eta (\varphi _{2}),...,\eta (\varphi
_{n}))d\mu (\eta )  \label{eq2.23}
\end{equation}%
is sequentially continuous in $\varphi _{j}\in F$, $j=\overline{1,m}$. (It
is well known \cite{GV,Go} that in countably normed spaces the properties of
sequential and ordinary continuity are equivalent).
\end{definition}

\begin{definition}
\label{def2.7} A cylindrical measure $\mu $ is countably additive if and
only if for any cylinder set $X=\coprod_{j\in \mathbb{Z}_{+}}X_{j}$, which
is the union of countably many mutually disjoints cylinder sets $%
X_{j}\subset F^{\prime }$, $j\in \mathbb{Z}_{+}$, $\mu (X)=\sum_{j\in 
\mathbb{Z}_{+}}\mu (X_{j})$.
\end{definition}

The following propositions hold.

\begin{proposition}
\label{pr2.8} A countably additive cylindrical measure $\mu $ can be
extended to a countably additive measure on the $\sigma -$ algebra,
generated by the cylinder sets with Borel base. Such a measure will also  be
called a cylindrical measure.
\end{proposition}

\begin{proposition}
\label{pr2.9} Let $F$ be a nuclear space. Then any cylindrical measure $\mu$
on $F^{\prime }$, satisfying the continuity condition, is countably additive.
\end{proposition}

\begin{definition}
\label{def2.10} Let $\mu$ be a cylindrical measure in $F^{\prime }$. The
Fourier transform of $\mu$ is the nonlinear functional 
\begin{equation}  \label{eq2.24}
\mathcal{L}(\varphi):=\int_{F^{\prime }}\exp[i\eta(\varphi)]d\mu(\eta).
\end{equation}
\end{definition}

\begin{definition}
\label{def2.11} The nonlinear functional $\mathcal{L}:F\longrightarrow%
\mathbb{C}$ on $F$, defined by (\ref{eq2.24}), is called positive definite,
if and only if for all $f_{j}\in F$ and $\lambda_{j}\in\mathbb{C}$, $j=%
\overline{1,n}$, the condition 
\begin{equation}  \label{eq2.25}
\sum_{j,k=1}^{n}\bar{\lambda}_{j}\mathcal{L}(f_{k}-f_{j})\lambda_{k}\geq0
\end{equation}
holds for any $n\in\mathbb{Z}_{+}$.
\end{definition}

\begin{proposition}
\label{pr2.12} The functional $\mathcal{L}:F\longrightarrow\mathbb{C}$ on $F$%
, defined by (\ref{eq2.24}), is the Fourier transform of a cylindrical
measure on $F^{\prime }$, if and only if it is positive definite,
sequentially continuous and satisfying the condition $\mathcal{L}(0)=1$.
\end{proposition}

Suppose now that we have a continuous unitary representation of the unitary
family $\mathcal{U}$ in a Hilbert space $\Phi $ with a cyclic vector $%
|\Omega \rangle \in \Phi $. Then we can put 
\begin{equation}
\mathcal{L}(\varphi ):=\langle \Omega |U(\varphi )|\Omega \rangle 
\label{eq2.26}
\end{equation}%
for any $\varphi \in F:=\mathcal{S}$, being the Schwartz space on $\mathbb{R}%
^{m}$, and observe that functional (\ref{eq2.26}) is continuous on $F$ owing
to the continuity of the representation. Therefore, this functional is the
generalized Fourier transform of a cylindrical measure $\mu $ on $\mathcal{S}%
^{\prime }:$ 
\begin{equation}
\langle \Omega |U(\varphi )|\Omega \rangle =\int_{\mathcal{S}^{\prime }}\exp
[i\eta (\varphi )]d\mu (\eta ).  \label{eq2.27}
\end{equation}%
From the spectral point of view, based on Theorem \ref{th2.1}, there is an
isomorphism between the Hilbert spaces $\Phi $ and $L_{2}^{(\mu )}(\mathcal{S%
}^{\prime };\mathbb{C})$, defined by $|\Omega \rangle \longrightarrow \Omega
(\eta )=1$ and $U(\varphi )|\Omega \rangle \longrightarrow \exp [i\eta
(\varphi )]$ and next extended by linearity upon the whole Hilbert space $%
\Phi $.

In the case of the non-cyclic case there exists a finite or countably
infinite family of measures $\left\{ \mu _{k}:k\in \mathbb{Z}_{+}\right\} $
on $\mathcal{S}^{\prime }$, with ${\Phi }\simeq \oplus _{k\in \mathbb{Z}%
_{+}}L_{2}^{(\mu _{k})}(\mathcal{S}^{\prime };\mathbb{C})$ and the unitary
operator $U(\varphi ):{\Phi }\longrightarrow {\Phi }$ for any $\varphi \in 
\mathcal{S}^{\prime }$ corresponds in all $L_{2}^{(\mu _{k})}(\mathcal{S}%
^{\prime };\mathbb{C})$, $k\in \mathbb{Z}_{+}$, to $\exp [i\eta (\varphi )]$%
. This means that there exists a single cylindrical measure $\mu $ on $%
\mathcal{S}^{\prime }$ and a $\mu -$measurable field of Hilbert spaces ${%
\Phi }_{\eta }$ on $\mathcal{S}^{\prime }$, such that 
\begin{equation}
{\Phi }\simeq \int_{\mathcal{S}^{\prime }}^{\oplus }{\Phi }_{\eta }d\mu
(\eta ),  \label{eq2.28}
\end{equation}%
with $U(\varphi ):{\Phi }\longrightarrow {\Phi }$, corresponding \cite{GV}
to the operator of multiplication by $\exp [i\eta (\varphi )]$ for any $%
\varphi \in \mathcal{S}$ and $\eta \in \mathcal{S}^{\prime }$. Thereby,
having constructed the nonlinear functional (\ref{eq2.24}) in an exact
analytical form, one can retrieve the representation of the unitary family $%
\mathcal{U}$ in the corresponding Hilbert space ${\Phi }$ of the Fock type,
making use of the suitable factorization (\ref{eq2.20}) as follows: ${\Phi }%
=\oplus _{n\in \mathbb{Z}_{+}}{\Phi }_{n}$, where 
\begin{equation}
{\Phi }_{n}=\underset{f_{n}\in L_{2,s}(\mathbb{(}{R}^{m})^{n};\mathbb{C})}{%
span}\left\{ \prod_{j=\overline{1,n}}a^{+}(x_{j})|\Omega \rangle \right\} ,
\label{eq2.29}
\end{equation}%
for all $n\in \mathbb{Z}_{+}$. The cyclic vector $|\Omega \rangle \in {\Phi }
$ can be, in particular, obtained as the ground state vector of some
unbounded self-adjoint positive definite Hamilton operator $\mathbb{H}:{\Phi 
}\longrightarrow {\Phi }$, commuting with the self-adjoint particles number
operator 
\begin{equation}
\mathbb{N}:=\int_{\mathbb{R}^{m}}\rho (x)dx,  \label{eq2.30}
\end{equation}%
that is $[\mathbb{H},\mathbb{N}]=0$. Moreover, the conditions 
\begin{equation}
\mathbb{H}|\Omega \rangle =0  \label{eq2.31}
\end{equation}%
and 
\begin{equation}
\inf_{g\in dom\mathbb{H}}\langle g,\mathbb{H}g\rangle =\langle \Omega |%
\mathbb{H}|\Omega \rangle =0  \label{eq2.32}
\end{equation}%
hold for the operator $\mathbb{H}:{\Phi }\longrightarrow {\Phi }$, where $%
dom \mathbb{H}$ denotes its domain of definition.

To find the functional (\ref{eq2.26}), which is called the generating
Bogolubov type functional for moment distribution functions 
\begin{equation}
F_{n}(x_{1},x_{2},...,x_{n}):=\langle \Omega |:\rho (x_{1})\rho
(x_{2})...\rho (x_{n}):|\Omega \rangle ,  \label{eq2.33}
\end{equation}%
where $x_{j}\in \mathbb{R}^{m}$, $j=\overline{1,n}$, and the normal ordering
operation $:\cdot :$ is defined as 
\begin{equation}
:\rho (x_{1})\rho (x_{2})...\rho (x_{n}):\;=\prod_{j=1}^{n}\left( \rho
(x_{j})-\sum_{k=1}^{j}\delta (x_{j}-x_{k})\right) ,  \label{eq2.34}
\end{equation}%
it is convenient to choose the Hamilton operator $\mathbb{H}:{\Phi }%
\longrightarrow {\Phi }$ in the following \cite{GGPS,Go,BP} algebraic form: 
\begin{equation}
\mathbb{H}:=\frac{1}{2}\int_{\mathbb{R}^{m}}K^{+}(x)\rho
^{-1}(x)K(x)dx+V(\rho ),  \label{eq2.35}
\end{equation}%
being equivalent in the Hilbert space $\Phi $ to the positive definite
operator expression 
\begin{equation}
\mathbb{H}:=\frac{1}{2}\int_{\mathbb{R}^{m}}(K^{+}(x)-A(x;\rho ))\rho
^{-1}(x)(K(x)-A(x;\rho ))dx,  \label{eq2.35a}
\end{equation}%
where $A(x;\rho ):\Phi \rightarrow \Phi ,$ $\ x\in \mathbb{R}^{m},$ is some
specially chosen linear self-adjoint operator. The "potential" operator $%
V(\rho ):{\Phi }\longrightarrow {\Phi }$ is, in general, a polynomial (or
analytical) functional of the density operator $\rho (x):{\Phi }%
\longrightarrow {\Phi }$ and the operator is given as 
\begin{equation}
K(x):=\nabla _{x}\rho (x)/2+iJ(x),  \label{eq2.36}
\end{equation}%
where the self-adjoint "current" operator $J(x):{\Phi }\longrightarrow {\Phi 
}$ can be defined (but non-uniquely) from the equality 
\begin{equation}
\partial \rho /\partial t=\frac{1}{i}[\mathbb{H},\rho (x)]=-<\nabla
_{x}\cdot J(x)>_{,}  \label{eq2.37}
\end{equation}%
holding for all $x\in \mathbb{R}^{m}$. Such an operator $J(x):{\Phi }%
\longrightarrow {\Phi }$, $x\in \mathbb{R}^{m}$ can exist owing to the
commutation condition $[\mathbb{H},\mathbb{N}]=0$, giving rise to the
continuity relationship (\ref{eq2.37}), if taking into account that supports 
$supp\;\rho $ of the density operator $\rho (x):{\Phi }\longrightarrow {\Phi 
}$, $x\in \mathbb{R}^{m}$, can be chosen arbitrarily owing to the
independence of (\ref{eq2.37}) on the potential operator $V(\rho ):{\Phi }%
\longrightarrow {\Phi },$ but its strict dependence on the corresponding
representation (\ref{eq2.28}). Denote also that representation (\ref{eq2.35a}%
) holds only under the condition that there exists such a self-adjoint
operator $A(x;\rho ):{\Phi }\longrightarrow {\Phi }$, $x\in \mathbb{R}^{m}$,
that 
\begin{equation}
K(x)|\Omega \rangle =A(x;\rho )|\Omega \rangle   \label{eq2.38}
\end{equation}%
for all ground states $|\Omega \rangle \in {\Phi }$, correspond to suitably
chosen potential operators $V(\rho ):{\Phi }\longrightarrow {\Phi }$.

The self-adjointness of the operator $A(x;\rho ):{\Phi }\longrightarrow {%
\Phi }$, $x\in \mathbb{R}^{m}$, can be stated following schemes from works 
\cite{GGPS,BP}, under the additional condition of the existence of such a
linear anti-unitary mapping $T:{\Phi }\longrightarrow {\Phi }$ that the
following invariance conditions hold: 
\begin{equation}
T\rho (x)T^{-1}=\rho (x),\qquad T\;J(x)\;T^{-1}=-J(x),\qquad T|\Omega
\rangle =|\Omega \rangle  \label{eq2.39}
\end{equation}%
for any $x\in \mathbb{R}^{m}$. Thereby, owing to conditions (\ref{eq2.39}),
the following expressions 
\begin{equation}
K^{\ast }(x )|\Omega \rangle =A(x;\rho )|\Omega \rangle =K(x)|\Omega \rangle
\label{eq2.40}
\end{equation}%
hold for any $x\in \mathbb{R}^{m}$, giving rise to the self-adjointness of
the operator $A(x;\rho ):{\Phi }\longrightarrow {\Phi }$, $x\in \mathbb{R}%
^{m}$.

Based now on the construction above one easily deduces from expression (\ref%
{eq2.37}) that the generating Bogolubov type functional (\ref{eq2.26}) obeys
for all $x\in \mathbb{R}^{m}$ the following functional-differential
equation: 
\begin{equation}
\left[ \nabla _{x}-i\nabla _{x}\varphi \right] \frac{1}{2i}\frac{\delta 
\mathcal{L}(\varphi )}{\delta \varphi (x)}=A\left( x;\frac{1}{i}\frac{\delta 
}{\delta \varphi }\right) \mathcal{L}(\varphi ),  \label{eq2.41}
\end{equation}%
whose solutions should satisfy the Fourier transform representation (\ref%
{eq2.27}). In particular, a wide class of special so-called Poissonian white
noise type solutions to the functional-differential equation (\ref{eq2.41})
was obtained in \cite{GGPS,BP} by means of functional-operator methods in
the following generalized form: 
\begin{equation}
\mathcal{L}(\varphi )=\exp \left\{ A\left( \frac{1}{i}\frac{\delta }{\delta
\varphi }\right) \right\} \exp \left( \bar{\rho}\int_{\mathbb{R}^{m}}\{\exp
[i\varphi (x)]-1\}dx\right) ,  \label{eq2.41a}
\end{equation}%
where $\bar{\rho}:=\langle \Omega |\rho |\Omega \rangle \in \mathbb{R}_{+}$
is a Poisson distribution density parameter.

Consider now the case, when the basic Fock space ${\Phi }=\otimes _{j=1}^{s}{%
\Phi }^{(j)}$, where ${\Phi }^{(j)}$, $j=\overline{1,s}$, are Fock spaces
corresponding to the different types of independent cyclic vectors $|\Omega
_{j}\rangle \in {\Phi }^{(j)}$, $j=\overline{1,s}.$ This, in particular,
means that the suitably constructed creation and annihilation operators $%
a_{j}(x),a_{k}^{+}(y):{\Phi }\longrightarrow {\Phi }$, $j,k=\overline{1,s}$,
satisfy the following commutation relations: 
\begin{equation}
\begin{array}{c}
\lbrack a_{j}(x),a_{k}(y)]=0, \\[5pt] 
\lbrack a_{j}(x),a_{k}^{+}(y)]=\delta _{jk}\delta (x-y)%
\end{array}
\label{eq2.42}
\end{equation}%
for any $x,y\in \mathbb{R}^{m}$.

\begin{definition}
\label{def2.13} A vector $|u\rangle \in {\Phi }$, $x\in \mathbb{R}^{m}$, is
called coherent with respect to a mapping $u\in L_{2}(\mathbb{R}^{m};\mathbb{%
R}^{s}):=M,$ if it satisfies the eigenfunction condition 
\begin{equation}
a_{j}(x)|u\rangle =u_{j}(x)|u\rangle  \label{eq2.43}
\end{equation}%
for each $j=\overline{1,s}$ and all $x\in \mathbb{R}^{m}$.
\end{definition}

It is easy to check that the coherent vectors $|u\rangle \in {\Phi }$ exist.
Really, the following vector expression 
\begin{equation}
|u\rangle :=\exp \{(u,a^{+})\}|\Omega \rangle ,  \label{eq2.44}
\end{equation}%
where $(.,.)$ is the standard scalar product in the Hilbert space $M$,
satisfies the defining condition (\ref{eq2.43}), and moreover, the norm 
\begin{equation}
\Vert u\Vert _{{\Phi }}:=\langle u|u\rangle ^{1/2}=\exp (\frac{1}{2}\Vert
u\Vert ^{2})<\infty ,  \label{eq2.45}
\end{equation}%
since $u\in M$ and its norm $\Vert u\Vert :=(u,u)^{1/2}$ is bounded.

\section{The Fock space embedding method, nonlinear dynamical systems and
their complete linearization}

Consider any function $u\in M:=L_{2}(\mathbb{R}^{m};\mathbb{R}^{s})$ and
observe that the Fock space embedding mapping 
\begin{equation}
\xi :M\ni u\longrightarrow |u\rangle \in {\Phi },  \label{eq3.1}
\end{equation}%
defined by means of the coherent vector expression (\ref{eq2.44}) realizes a
smooth isomorphism between Hilbert spaces $M$ and ${\Phi }$. The inverse
mapping $\xi ^{-1}:{\Phi }\longrightarrow M$ is given by the following exact
expression: 
\begin{equation}
u(x)=\langle \Omega |a(x)|u\rangle ,  \label{eq3.2}
\end{equation}%
holding for almost all $x\in \mathbb{R}^{m}$. Owing to condition (\ref%
{eq2.45}), one finds from (\ref{eq3.2}) that, the corresponding function $%
u\in M$.

In the Hilbert space $M,$ let now define a nonlinear dynamical system (which
can, in general, be  non-autonomous) in partial derivatives 
\begin{equation}
du/dt=K[u],  \label{eq3.3}
\end{equation}%
where $t\in \mathbb{R}_{+}$ is the corresponding evolution parameter, $%
[u]:=(t,x;u,u_{x},u_{xx},...,u_{rx}),r\in \mathbb{Z}_{+}$, and a mapping $%
K:M\longrightarrow T(M)$ is Frechet smooth. Assume also that the Cauchy
problem 
\begin{equation}
u|_{t=+0}=u_{0}  \label{eq3.4}
\end{equation}%
is solvable for any $u_{0}\in M$ in an interval $[0,T)\subset \mathbb{R}%
_{+}^{1}$ for some $T>0$. Thereby, the smooth evolution mapping is defined 
\begin{equation}
T_{t}:M\ni u_{0}\longrightarrow u(t|u_{0})\in M,  \label{eq3.5}
\end{equation}%
for all $t\in \lbrack 0,T)$.

It is now natural to consider the following commuting diagram 
\begin{equation}
\begin{array}{ccc}
M & \overset{\xi }{\longrightarrow } & \Phi \\ 
T_{t}\downarrow &  & \downarrow {\mathbb{T}_{t}} \\ 
M & \overset{\xi }{\longrightarrow } & \Phi,%
\end{array}
\label{eq3.6}
\end{equation}%
where the mapping $\mathbb{T}_{t}:{\Phi }\longrightarrow {\Phi }$, $t\in
\lbrack 0,T)$, is defined from the conjugation relationship 
\begin{equation}
\xi \circ T_{t}=\mathbb{T}_{t}.\circ \xi  \label{eq3.7}
\end{equation}

Now take coherent vector $|u_{0}\rangle \in {\Phi }$, corresponding to $%
u_{0}\in M,$ and construct the vector 
\begin{equation}
|u\rangle :=\mathbb{T}_{t}\cdot |u_{0}\rangle   \label{eq3.8}
\end{equation}%
for all $t\in \lbrack 0,T)$. Since vector (\ref{eq3.8}) is, by construction,
coherent, that is 
\begin{equation}
a_{j}(x)|u\rangle :=u_{j}(x,t|u_{0})|u\rangle   \label{eq3.9}
\end{equation}%
for each $j=\overline{1,s}$, $t\in \lbrack 0,T)$ and almost all $x\in 
\mathbb{R}^{m}$, owing to the smoothness of the mapping $\xi
:M\longrightarrow {\Phi }$ with respect to the corresponding norms in the
Hilbert spaces $M$ and $\Phi ,$ we derive that coherent vector (\ref{eq3.8})
is differentiable with respect to the evolution parameter $t\in \lbrack 0,T)$%
. Thus, one can easily find \cite{KS,Ko} that 
\begin{equation}
\frac{d}{dt}|u\rangle =\hat{K}[a^{+},a]|u\rangle ,  \label{eq3.10}
\end{equation}
where 
\begin{equation}
|u\rangle |_{t=+0}=|u_{0}\rangle   \label{eq3.11}
\end{equation}%
and a mapping $\hat{K}[a^{+},a]:{\Phi }\longrightarrow {\Phi }$ is defined
by the exact analytical expression 
\begin{equation}
\hat{K}[a^{+},a]:=(a^{+},K[a]).  \label{eq3.12}
\end{equation}

As a result of the consideration above we obtain the following theorem.

\begin{theorem}
\label{th3.1} Any smooth nonlinear dynamical system (\ref{eq3.3}) in Hilbert
space $M:=L_{2}(\mathbb{R}^{m};\mathbb{R}^{s})$ is representable by means of
the Fock space embedding isomorphism $\xi :M\longrightarrow {\Phi }$ in the
completely linear form (\ref{eq3.10}).
\end{theorem}

We now make some comments concerning the solution to the linear equation (%
\ref{eq3.10}) under the Cauchy condition (\ref{eq3.11}). Since any vector $%
|u\rangle \in {\Phi }$ allows the series representation 
\begin{equation}
\begin{array}{l}
|u\rangle =\underset{n:=\sum_{j=1}^{s}n_{j}\in \mathbb{Z}_{+}}{\bigoplus }%
\frac{1}{(n_{1}!n_{2}!...n_{s}!)^{1/2}}\int_{(\mathbb{R}%
^{m})^{n}}f_{n_{1}n_{2}...n_{s}}^{(n)}(x_{1}^{(1)},x_{2}^{(1)},...,x_{{n}%
_{1}}^{(1)}; \\[10pt]
\qquad x_{1}^{(2)},x_{2}^{(2)},...,x_{{n}%
_{2}}^{(2)};...;x_{1}^{(s)},x_{2}^{(s)},...,x_{{n}_{s}}^{(s)})%
\prod_{j=1}^{s}\left(
\prod_{k=1}^{n_{j}}dx_{k}^{(j)}a_{j}^{+}(x_{k}^{(j)})\right) |\Omega \rangle
,%
\end{array}
\label{eq3.13}
\end{equation}%
where for any $n=\sum_{j=1}^{s}n_{j}\in \mathbb{Z}_{+}$ functions 
\begin{equation}
f_{n_{1}n_{2}...n_{s}}^{(n)}\in \bigotimes_{j=1}^{s}L_{2,s}((\mathbb{R}%
^{m})^{n_{j}};\mathbb{C})\simeq L_{2,s}(\mathbb{R}^{mn_{1}}\times \mathbb{R}%
^{mn_{2}}\times ...\mathbb{R}^{mn_{s}};\mathbb{C}),  \label{eq3.14}
\end{equation}%
and the norm 
\begin{equation}
\Vert u\Vert _{{\Phi }}^{2}=\sum_{n=\sum_{j=1}^{s}n_{j}}\Vert
f_{n_{1}n_{2}...n_{s}}^{(n)}\Vert _{2}^{2}=\exp (\Vert u\Vert ^{2}).
\label{eq3.15}
\end{equation}%
By substituting (\ref{eq3.13}) into equation (\ref{eq3.10}),  reduces (\ref%
{eq3.10}) to an infinite recurrent set of linear evolution equations in
partial derivatives on coefficient functions (\ref{eq3.14}). The latter can
often be solved \cite{Ko} step by step analytically in exact form, thereby,
making it possible to obtain, owing to representation (\ref{eq3.2}), the
exact solution $u\in M$ to the Cauchy problem (\ref{eq3.4}) for our
nonlinear dynamical system in partial derivatives (\ref{eq3.3}).

\begin{remark}
\label{re3.2} Concerning some applications of nonlinear dynamical systems
like (\ref{eq3.1}) in mathematical physics problems, it is very important to
construct their so called conservation laws or smooth invariant functionals $%
\gamma :M\longrightarrow \mathbb{R}$ on $M$. Making use of the quantum
mathematics technique described above one can suggest an effective algorithm
for constructing these conservation laws in exact form.
\end{remark}

Indeed, consider a vector $|\gamma \rangle \in {\Phi }$, satisfying the
linear equation: 
\begin{equation}
\frac{\partial }{\partial t}|\gamma \rangle +\hat{K}^{\ast }[a^{+},a]|\gamma
\rangle =0.  \label{eq3.16}
\end{equation}

Then, the following proposition \cite{Ko} holds.

\begin{proposition}
\label{pr3.3} The functional 
\begin{equation}
\gamma :=\langle u|\gamma \rangle  \label{eq3.17}
\end{equation}%
is a conservation law for dynamical system (\ref{eq3.1}), that is 
\begin{equation}
d\gamma /dt|_{K}=0  \label{eq3.18}
\end{equation}%
along any orbit of the evolution mapping (\ref{eq3.5}).
\end{proposition}

\section{Conclusion}

Within the scope of this work we have described the main mathematical
preliminaries and properties of the quantum mathematics techniques suitable
for analytical studying of the important linearization problem for a wide
class of nonlinear dynamical systems in partial derivatives in Hilbert
spaces. This problem was analyzed in much detail using the Gelfand-Vilenkin
representation theory \cite{GV} of infinite dimensional groups and the
Goldin-Menikoff-Sharp theory \cite{GGPS,Go,GMS} of generating Bogolubov type
functionals, classifying these representations. The related problem of
constructing Fock type space representations and retrieving their
creation-annihilation generating structure still needs a deeper
investigation within the approach devised. Here we mention only that some
aspects of this problem within the so-called Poissonian White noise analysis
were studied in a series of works \cite{Be,AKS,KSWY,LRS}, based on some
generalizations of the Delsarte type characters technique. It is also
necessary to mention the related results obtained in \cite{KS1,Ko,KS},
devoted to the application of the Fock space embedding method to finding
conservation laws and the so called recursion operators for the well known
Korteweg-de Vries type nonlinear dynamical systems. We plan to devote our
next investigations to concerning some important applications of the methods
devised in the work to concrete dynamical systems.

\section*{Acknowledgments}

Two of the authors (N.B. and A.P.) are cordially thankful to the Abdus Salam
International Centre for Theoretical Physics in Trieste, Italy, for the
hospitality during their research 2007-scholarships. A.P. is also thankful
to Profs. Y. Kondratyev and Y. Kozicki for interesting discussions of the
results during the Third International Conference on Infinite Dimensional
Systems, held June 23-28, 2007 in Kazimierz Dolny, Poland, and to Profs. O.
Celebi, K. Zheltukhin, G. Barsegian, O. Zhdanov and V. Golubeva during the
ISAAC-2007 International Conference, held 13-18 August 2007 in Ankara,
Turkey.


\newpage
\begin{thebibliography}{99}
\bibitem{AKS} Albeverio S., Kondratiev Y.G. and Streit L. How to generalize
white noice analysis to non-gaussian measures. Preprint Bi Bo S, Bielefeld,
1992.

\bibitem{Be} Berezansky Y.M. A generalization of white noice analysis by
means of theory of hypergroups. Reports on Math. Phys., 38, N.3 (1996), pp.
289-300.

\bibitem{BK} Berezansky Y.M. and Kondratiev Y.G. Spectral methods in
infinite dimensional analysis, v.1 and 2, Kluwer, 1995.

\bibitem{BB} Bogolubov N.N. and Bogolubov N.N. (jr.) Introduction into
quantum statistical mechanics. World Scientific, NJ, 1986, 384P.

\bibitem{BP} Bogoliubov N.N. (Jr.), Prykarpatsky A.K. Quantum method of
generating Bogolubov functionals in statistical physics: current Lie
algebras, their representations and functional equations. Physics of
Elementary Particles and Atomique Nucleus, v.17, N.4 (1986), pp. 791-827 (in
Russian).

\bibitem{Ber} Berezin F.A. The second quantization method. \ Nauka
Puplisher, Moscow, 1986 (in Russian).

\bibitem{Bo} Bogoliubov N.N. Collected works, v.2, Naukova Dumka, Kiev, 1960
(in Russian).

\bibitem{Dir} Dirac P.A.M. The principles of quantum mechanics. Oxford
University Press, 1932.

\bibitem{Di} Dixmier J. C*-algebras. Amsterdam, North-Holland, 1982.

\bibitem{Fa} Faddeev L.D. , Sklyanin E.K. Quantum mechanical approach to
completely integrable field theories. Proceed. of the USSR Academy of
Sciences (DAN), 243 (1978), pp. 1430-1433 (in Russian).

\bibitem{Fe} Feynman R. Quantum mechanical computers. Found. Physics, 16
(1986), pp. 507-531.

\bibitem{Fe1} Feynman R. Simulating physics with computers. Intern. Journal
of Theor. Physics, 21 (1982), pp. 467-488.

\bibitem{Fo} Fock V.A. \ Konfigurationsraum und zweite Quantelung.
Zeischrift Phys., Bd. 75 (1932), pp. 622-647.

\bibitem{GV} Gelfand I. and Vilenkin N. Generalized functions, 4, Academic
Press, New York, 1964.

\bibitem{Go} Goldin G.A. Nonrelativistic current algebras as unitary
representations of groups. Journal of Mathem. Physics, 12(3), 1971, pp.
462-487.

\bibitem{GGPS} Goldin G.A., Grodnik J., Powers R.T. and Sharp D.
Nonrelativistic current algebra in the N/V limit. J. Math. Phys., 15,
(1974), pp. 88-100.

\bibitem{GMS} Goldin G.A., Menikoff R. and Sharp F.H. Diffeomorphism groups,
gauge groups, and quantum theory. Phys. Rev. Lett. 51 (1983), pp. 2246-2249.

\bibitem{Gr} Grover L.K. Quantum mechanics helps in searching for a needle
in a haystack. Phys. Rev.\ Lett., 79 (1997), pp. 325-328.

\bibitem{JQ} Jaffe A. and Quinn F. Theoretical mathematics: toward a
cultural synthesis of mathematics and theoretical physics. Bull. Amer. Math.
Soc., 29 (1993), pp. 1-13 Zeischrift Phys., Bd. 75 (1932), pp. 622-647.

\bibitem{De} Deutsch D. Quantum theory, the Church-Turing principle and the
universal quantum computer. Proc. Roy. Soc. (London), A400 (1985), pp.
97-117.

\bibitem{Dr} Drinfeld V.G. Quantum groups. Proceed. of the Int. Congress of
Mathematicians, 1986, pp. 798-820.

\bibitem{Do} Donaldson S.K. An application of gauge theory to four
dimansional topology. J. Diff. Geom., 17 (1982), pp. 279-315.

\bibitem{KS1} Kowalski K. and Steeb W.-H. Symmetries and first integrals for
nonlinear dynamical systems: Hilbert space approach. I and II. Progress of
Theoretical Physics, 85, N.4 (1991), pp. 713-722 and 85, N4 (1991), pp.
975-983.

\bibitem{KSWY} Kondratiev Y.G., Streit L., Westerkamp W. and Yan J.-A.
Generalized functions in infinite dimensional analysis. II AS preprint, 1995.

\bibitem{Ko} Kowalski K. \ Methods of Hilbert spaces in the theory of
nonlinear dynamical systems. World Scientific, 1994.

\bibitem{KS} Kowalski K. and Steeb W.-H. Non linear dynamical systems and
Carleman linearization. World Scientific, 1991.

\bibitem{LRS} Lytvynov E.W., Rebenko A.L. and Shchepaniuk G.V. Wick calculus
on spaces of generalized functions compound Poisson white noise. Reports on
Math. Phys., 39, N.2 (1997), pp. 219-247.

\bibitem{Ma} Manin Yu.I. Computable and uncomputable. Moscow, Sov. Radio,
1980 (in Russian).

\bibitem{Ma1} Manin Yu.I. Classical computation, quantum computation and P.
Shor's factoraizing algorithm. Proceed. of the Bourbaki Seminar, 1999.

\bibitem{Ne} Neumann J. von. Mathematische Grundlagen der Quanten Mechanik.
J. springer, Berlin, 1932.

\bibitem{PTB} Prykarpatsky A.K., Taneri U. and Bogolubov N.N. (jr.) \
Quantum field theory and application to quantum nonlinear optics. World
Scientific, NY, 2002.

\bibitem{Sh} Shor P.W. Polynomial time algorithms for prime factorization
and discrete logarithms on a quantum computer. SIAM Journ. Comput., 26 N.5
(1997), pp. 1484-1509.

\bibitem{We} Weyl H. The Theory of Groups and Quantum Mechanics. Dover, New
York, 1931.

\bibitem{Wi} Witten E. Nonabelian bozonization in two dimensions. Commun.
Mathem. Physics, 92 (1984), pp. 455-472.
\end{thebibliography}
\end{document}